\documentclass[aps,prl,twocolumn,superscriptaddress]{revtex4}
\usepackage{amsmath}
\usepackage{amssymb}
\usepackage{hyperref}
\usepackage{url}
\usepackage{graphicx}
\usepackage{dcolumn} 
\usepackage{bm}      
\usepackage{multirow}
\usepackage{epstopdf}
\usepackage{color}
\usepackage{gensymb}
\usepackage{textcomp}
\usepackage{chemformula} 

\begin{document}

\title{Random Fields from Quenched Disorder in an Archetype for Correlated Electrons: the Parallel Spin Stripe Phase of La$_{1.6-x}$Nd$_{0.4}$Sr$_x$CuO$_4$ at the 1/8 Anomaly}

\author{Q.~Chen}
\affiliation{%
 Department of Physics and Astronomy, McMaster University, Hamilton, Ontario, L8S 4M1, Canada
}%
\affiliation{
 Brockhouse Institute for Materials Research, Hamilton, Ontario, L8S 4M1, Canada
}%

\author{S. H.-Y. Huang}
\author{Q. Ma}
\author{E. M. Smith}
\author{H. Sharron}
\affiliation{%
 Department of Physics and Astronomy, McMaster University, Hamilton, Ontario, L8S 4M1, Canada
}%

\author{A. A. Aczel}
\author{W. Tian}
\affiliation{%
    Neutron Scattering Division, Oak Ridge National Laboratory, Oak Ridge, TN 37831, USA
}%

\author{B. D. Gaulin}%
\affiliation{%
 Department of Physics and Astronomy, McMaster University, Hamilton, Ontario, L8S 4M1, Canada
}%
\affiliation{
 Brockhouse Institute for Materials Research, Hamilton, Ontario, L8S 4M1, Canada
}%
\affiliation{
 Canadian Institute for Advanced Research, Toronto, Ontario M5G 1M1, Canada
}%

\date{\today}

\begin{abstract}
The parallel stripe phase is remarkable both in its own right, and in relation to the other phases it co-exists with.
Its inhomogeneous nature makes such states susceptible to random fields from quenched magnetic vacancies.
We argue this is the case by introducing low concentrations of nonmagnetic Zn impurities (0-10\%) into La$_{1.6-x}$Nd$_{0.4}$Sr$_x$CuO$_4$ (Nd-LSCO) with $x$ = 0.125 in single crystal form, well below the percolation threshold of $\sim$\,41\% for two-dimensional (2D) square lattice. 
Elastic neutron scattering measurements on these crystals show clear magnetic quasi-Bragg peaks at all Zn dopings. While all the Zn-doped crystals display order parameters that merge into each other and the background at $\sim$\,68\,K, the temperature dependence of the order parameter as a function of Zn concentration is drastically different.
This result is consistent with meandering charge stripes within the parallel stripe phase, which are pinned in the presence of quenched magnetic vacancies. In turn it implies vacancies that preferentially occupy sites within the charge stripes, and hence that can be very effective at disrupting superconductivity in Nd-LSCO ($x$ = 0.125), and, by extension, in all systems exhibiting parallel stripes.   
\end{abstract}

\maketitle

\begin{figure}[!hbp]
\includegraphics[width=5cm]{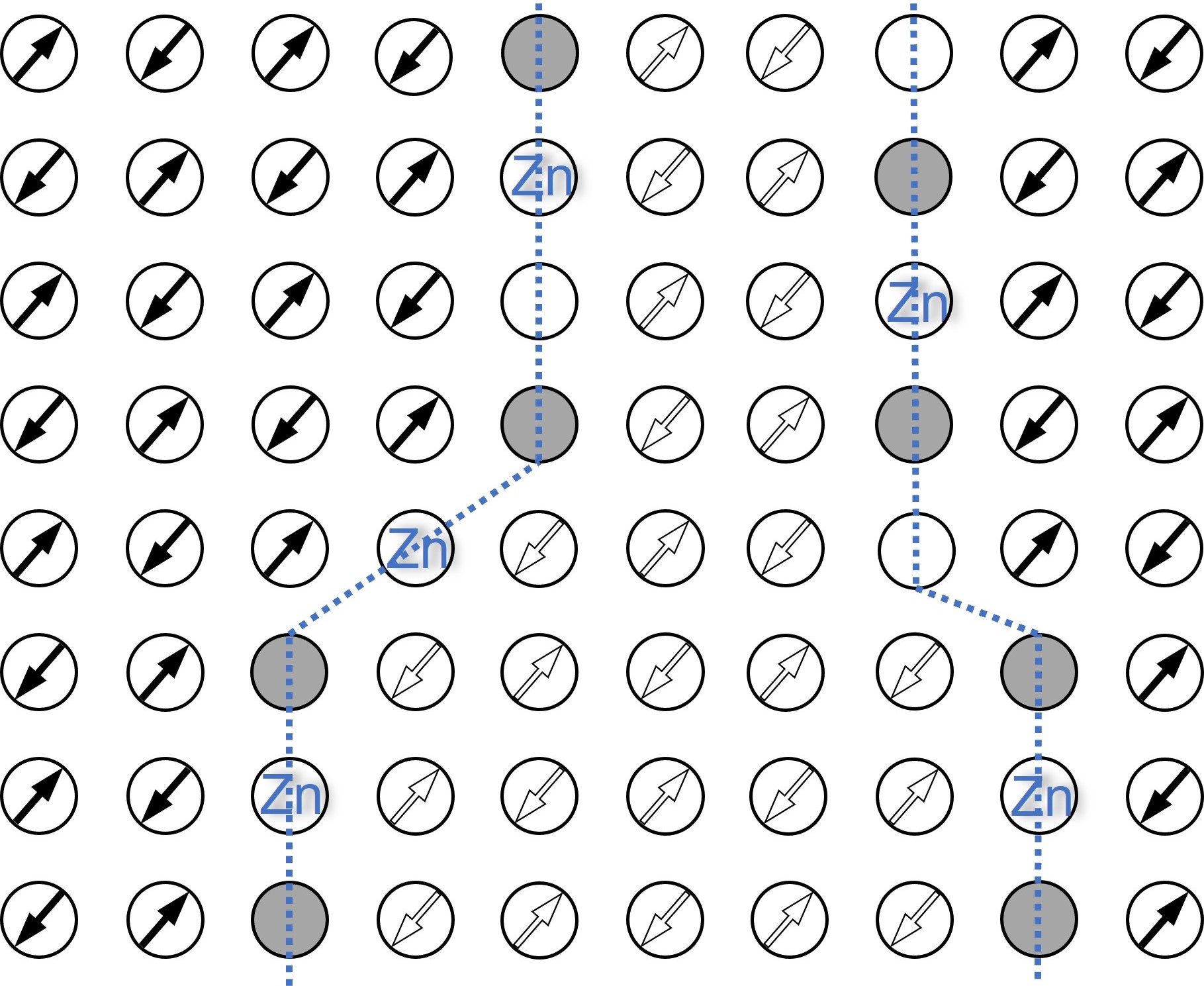}
\caption{A schematic drawing of parallel stripe order in the La214 cuprate with quenched nonmagnetic Zn  impurities replacing Cu in the CuO$_2$ plane. Only the Cu sites are shown. This 2D parallel stripe structure is an inhomogeneous arrangement of holes (filled grey circles), and spins (arrows) within a CuO$_2$ plane, while  the empty circles represent paramagnetic sites.  Shading of the arrowheads distinguishes the antiphase N\'eel domains.
}
\label{Fig1}
\end{figure}

Parallel stripe order and fluctuations have been proposed to underlie the mechanism for high-$T_\text{c}$ superconductivity in layered copper oxides. 
This unusual, inhomogeneous, intertwined spin and charge structure is described in terms of narrow strips of anti-phase N\'eel states that are separated from each other by charge stripes, as illustrated in Fig.~\ref{Fig1}.  
The parallel spin stripe phase in the cuprates is the strongest in Nd-LSCO at the ``1/8 anomaly" ($x$ = 0.125)\cite{95_tranquada,96_tranquada}, where its onset temperature was thought to maximize at $\sim$\,50\,K, and its superconducting $T_\text{c}$ is at a local minimum \cite{94_axe,19_michon,20_dragomir}.
The parallel charge stripes are observed at twice the incommensurate wavevectors of the spin stripes, and were thought to have a somewhat higher temperature onset \cite{95_tranquada}.  Together, these intertwined parallel spin and charge orders have been consistently interpreted in terms of the parallel stripe picture which possesses a remarkable {\it inhomogeneous} magnetic structure. This inhomogeneous magnetic structure has been observed to co-exist with superconductivity in Nd-LSCO, from $x$ = 0.05 to $x$ = 0.26 \cite{91_cheong, 92_mason,92_thurston,01_wakimoto,04_fujita,06_birgeneau,21_ma,22_ma}. However, whether stripes help or hinder superconductivity remains a matter of debate.

Other than the collinear stripe picture \cite{17_charlebois,90_schulz,01_fleck,03_kivelson,07_christensen}, the incommensurate magnetic peaks observed in the cuprates are also discussed in terms of spiral spin density wave (SDW) states caused by Fermi surface nesting \cite{89_shraiman,06_kotov,16_yamase,16_eberlein}. This inherently itinerant origin for a form of spiral magnetism does {\it not} require inhomogeneity. In principle, neutron crystallographic techniques should be able to distinguish between inhomogeneous stripe and homogeneous spiral SDW structures, but the complexity of the relevant spin structures make this problem difficult \cite{07_christensen}.  

The subject of this Letter is the sensitivity of the parallel stripe phase to quenched disorder. It is interesting precisely because the parallel stripe phase possesses an inhomogeneous magnetic structure, and therefore quenched disorder is expected to couple to this structure as a random field.
In fact, quenched disorder in the form of magnetic vacancies has been well studied in the parent compound, quasi-2D quantum antiferromagnet La$_2$CuO$_4$, by jointly substituting Zn and Mg on the Cu site \cite{02_vajk}. This neutron scattering work shows beautifully how quenched disorder in this related but {\it homogeneous} magnetic structure, a simple two sub-lattice N\'eel state, gives rise to the expectations of the 2D percolation theory - a percolation threshold of $\sim$\,41$\%$ \cite{00_newman}.

Inhomogeneous magnetic structures are relatively rare in nature, but can occur, for example, in the presence of geometrical frustration in insulators. One well studied example is that of the spin-1/2 Ising-like stacked triangular lattice antiferromagnet CsCoBr$_3$ \cite{02_mao}.
Here, an inhomogeneous, partially paramagnetic structure exists over an extended range of temperature. Neutron diffraction studies on single crystal CsCo$_{0.83}$Mg$_{0.17}$Br$_3$ show that quenched magnetic vacancies (Mg substituting for Co) severely disrupt the nature of the magnetic order parameter in this system, over the temperature range of the inhomogeneous, partially paramagnetic phase  \cite{04_duijn}.

We propose that very similar phenomena occurs in the parallel stripe phase of the cuprates.
The pinning of nonmagnetic charge stripes by quenched magnetic vacancies within the parallel stripe structure is qualitatively illustrated in Fig. \ref{Fig1}. This figure shows a Zn concentration of $\sim$ 6\%, which causes fluctuations in the parallel spin stripe width. 
The charge stripes follow the local pattern of quenched Zn impurities to minimize the magnetic exchange interaction energy of a Zn impurity, compared with when the impurity is located within the middle of a N\'eel domain.
Although wider charge stripes are possible \cite{13_tranquada,20_tranquada,21_tranquada}, they are abstracted to being a single Cu-site across in Fig. \ref{Fig1}. The key characteristic of the charge stripe is that the parallel spin stripes on either side of it are antiphase N\'eel states, $\pi$ out-of-phase relative to each other, thus giving rise to the incommensurate nature of the magnetic Bragg peaks.
In what follows we present experimental evidence for this picture, and discuss implications for the nature of superconducting pairing in this, and by extension, many phases of cuprate superconductors.

High-quality single crystals of Zn-doped Nd-LSCO ($x$ = 0.125) were grown using the travelling solvent floating zone technique \cite{03_yan,20_dragomir,22_ma2}. 
Magnetization was measured on single crystals using a Quantum Design MPMS superconducting quantum interference device (SQUID) magnetometer.
Elastic neutron scattering measurements were performed with the fixed-incident-energy triple-axis spectrometer HB-1A at Oak Ridge National Laboratory.

\begin{figure}[!htbp]
\includegraphics[width=8cm]{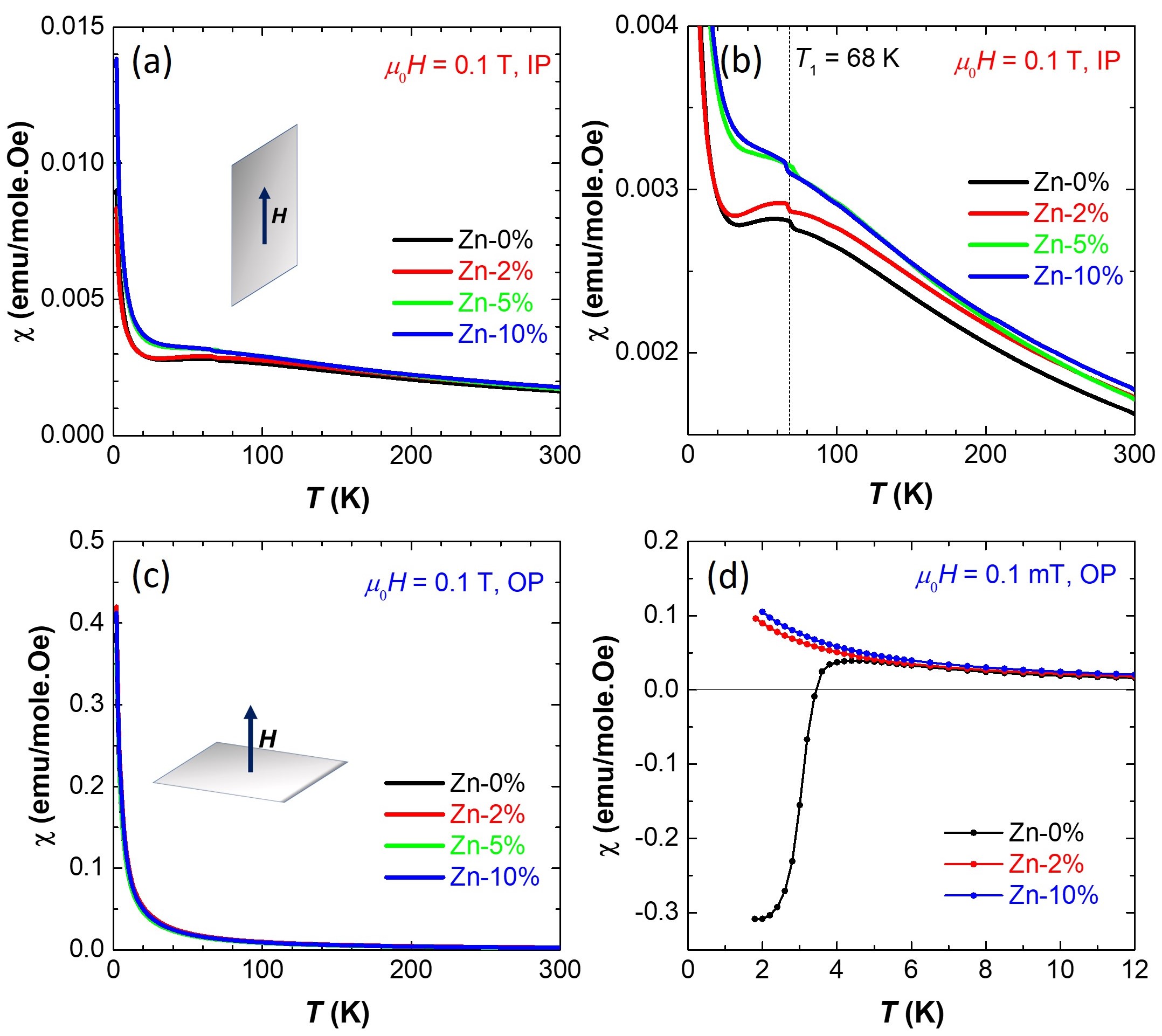}
\caption{Magnetic susceptibility measurements for Zn-doped single crystals of Nd-LSCO ($x$ = 0.125) with field (a-b) in-plane (IP) and (c-d) out-of-plane (OP). (a) and (b) show the same data on different scales. (c) and (d) show field OP measurements with 0.1 T and 0.1 mT, respectively. The insets in (a) and (c) illustrate the orientation of the CuO$_2$/$ab$ plane of the crystals relative to the external field.
}
\label{Fig2}
\end{figure}
 
Magnetic susceptibilities were measured to characterize the magnetic/charge order [Figs. \ref{Fig2}(a-c)] and superconducting [Fig.~\ref{Fig2}(d)] transitions. The zero-field-cooled (ZFC) warm-up, followed by field-cooled (FC) cool-down measurement protocols were employed. For clarity, only FC data are shown because the ZFC and FC data overlap, except for the Zn-0\% sample measured at 0.1\,mT, which shows the ZFC-FC bifurcation near the superconducting transition and is consistent with previous studies \cite{94_axe,20_dragomir}.

The $H \parallel ab$ (IP) measurements in Figs.~\ref{Fig2}(a-b) show a discontinuity at $T_\text{1}$\,$\sim$\,68\,K, coincident with the low-temperature-orthorhombic (LTO) to low-temperature-tetragonal (LTT) structural phase transition \cite{20_dragomir}. A broad peak, roughly centred at $T_\text{1}$, suggests the onset of strong antiferromagnetic fluctuations in the parallel spin stripe phases. While the field-IP data below $\sim$\,100\,K show systematic Zn-dependence, in contrast, the $H \parallel c$ (OP) data in Fig. \ref{Fig2}(c) are almost Zn-independent, likely due to the large random moments of Nd$^{3+}$ whose crystal field effects maintain their orientation along the $c$-axis.
In Fig.~\ref{Fig2}(d), the low-temperature measurements for the Zn-0\% sample show a sharp drop below $T_\text{c}$\,$\sim$\,3\,K, while no signs of superconducting transition are observed down to 1.8\,K for the other samples. This result, that 2\% of nonmagnetic Zn impurities suppress $T_\text{c}$ by a factor of at least 2, is consistent with previous studies of Zn-doping in LBCO, LSCO, YBCO and Nd-LSCO systems, where the suppression of $T_\text{c}$ by nonmagnetic Zn is described with the ``Swiss Cheese" model wherein charge carriers within an area of $\pi \xi_{ab}^2$ around each Zn are excluded from the superfluid, where $\xi_{ab}$ is the in-plane coherence length \cite{96_nachumi,96_fukuzumi,98_nakano,04_adachi,21_lozano,12_wenjs, 17_guguchia, 89_uemura}.  

\begin{figure*}[!htbp]
\includegraphics[width=17cm]{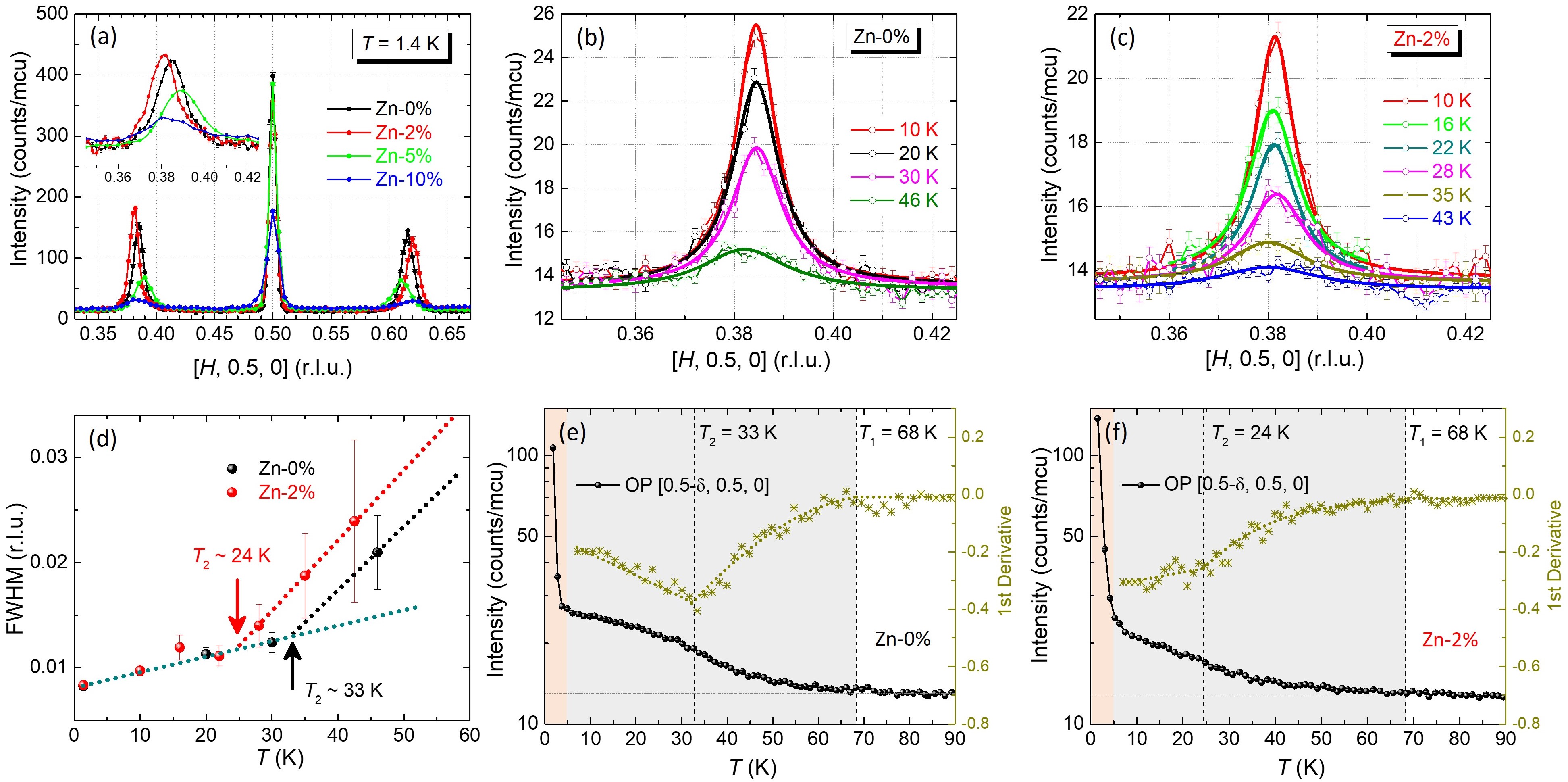}
\caption{(a) Elastic neutron scattering scans of the form [$H$, 0.5, 0] near $H$ = 0.5 at 1.4\,K. The inset shows the same intensities at $H$ = $0.5-\delta$ on log intensity scale. (b-c) Elastic $H$-scans through [$0.5- \delta$, 0.5, 0] for Zn-0\% and Zn-2\% samples at different temperatures. These data have been fit to Lorentzian line-shapes (solid lines). (d) FWHM extracted from these fits as a function of temperature. (e-f) Order parameter of the [$0.5-\delta$, 0.5, 0] ($\delta \approx$ 0.125) magnetic peak for Zn-0\% and Zn-2\%, respectively. Anomalies in the 1st derivative derived from these order parameters allow us to estimate $T_\text{2}$ $\sim$\,33(1)\,K and 24(2)\,K for Zn-0\% and Zn-2\%, respectively. The dashed lines are guides to the eyes. 
}
\label{Fig3}
\end{figure*}

Our elastic neutron scattering studies of the incommensurate magnetic Bragg peaks associated with the parallel spin stripe phase used $E_i=14.5$\,meV. Collimation of 40'-40'-40'-80' resulted in an energy resolution at the elastic line just over 1\,meV (FWHM). The single crystal samples were mounted so that the $(H K 0)$ peaks are in the scattering plane, where $HKL$ are defined in tetragonal notation with $a \simeq$ 3.78\,$\text{\AA}$ and $c \simeq$ 13.14\,$\text{\AA}$.

Elastic neutron scattering scans of the form [$H$, 0.5, 0] and [0.5, $K$, 0] near $H(K)$ = 0.5 were carried out for all four single crystals. Fig. \ref{Fig3}(a) shows the data measured at base temperature $T$ = 1.4\,K. Only $H$-scans are shown because they overlap with the corresponding $K$-scans. Incommensurate antiferromagnetic quasi-Bragg peaks are observed at [$0.5 \pm \delta$, 0.5, 0] and [0.5, $0.5 \pm \delta$, 0], with $\delta \approx$ 0.125 at all Zn dopings. Commensurate nuclear Bragg peaks are observed at [0.5, 0.5, 0], between the incommensurate antiferromagnetic quasi-Bragg peaks.  The temperature dependence of this commensurate nuclear scattering is sensitive to the structural LTO to LTT phase transition at $T_\text{1} \sim$ 68\,K. 

The inset of Fig. \ref{Fig3}(a) shows the comparison of the [$0.5 - \delta$, 0.5, 0] peaks on a log intensity scale. The peak intensity varies by a factor of $\sim$ 9 between the Zn-0\% and the Zn-10\% samples, and the high Zn-doped single crystals clearly exhibit quasi-Bragg peaks which are broader in the $(H K 0)$ plane. In addition, shifts in the peak position of up to 0.004 {\it r.l.u.} in $H$ can be seen.  These shifts are not systematic with Zn-doping, and are likely due to small, random variation in the Sr or hole concentration of single crystals by $\pm$\,0.004, assuming that the wavevector follows the Yamada relation $\delta \approx x$ \cite{98_yamada}. 

The order parameter of the incommensurate wavevector [$0.5 - \delta$, 0.5, 0] for the Zn-0\% and 2\% samples is shown in Fig. \ref{Fig3}(e) and (f), respectively. Our data for the pure sample (Zn-0\%) is very similar to that originally reported \cite{95_tranquada}, but with better counting statistics and a much increased temperature-point density, allowing a sensitive measurement of the form of the order parameter.  

At the lowest temperatures, below 5\,K, one sees a dramatic upturn in the order parameter, which was ascribed to the effect of coupling between the Nd$^{3+}$ moments randomly distributed over the La$^{3+}$ sites between the CuO$_2$ planes, to the Cu$^{2+}$ moments within the plane \cite{96_tranquada}.  Such coupling is known to develop three-dimensional (3D) correlations into the parallel spin stripe phase, which are absent above $\sim$ 5\,K.  This effect concentrates the elastic incommensurate magnetic scattering at [$0.5 - \delta$, 0.5, 0], as opposed to along the line [$0.5 - \delta$, 0.5, $L$].  This strong, quasi-3D parallel spin stripe order co-exists perfectly well with superconductivity below $T_\text{c}$. 

Above 5\,K, the order parameter for Zn-0\% shows typical behaviour, with downwards curvature approaching what appears to be a phase transition near $T_\text{2} \sim$ 33\,K. The quality of the order parameter data for the Zn-0\% and Zn-2\% samples in Fig. \ref{Fig3}(e) and (f) is sufficiently high that it can be fit to a polynomial expansion such that the first derivative of the order parameter can also be obtained as a function of temperature.  This is shown in Fig. \ref{Fig3}(e) and (f) on the right hand scale.  For Zn-0\% in Fig. \ref{Fig3}(e), a sharp change in slope is observed at $T_\text{2}$ = 33(1)\,K, where the order parameter changes from upwards curvature to downwards curvature.  A similar feature is observed at $T_\text{2}$ = 24(2)\,K for Zn-2\% in Fig. \ref{Fig3}(f).  Note that $T_\text{2} \sim$ 30\,K has been previously identified as the onset temperature for static magnetism in pure Nd-LSCO ($x$ = 0.125) by $\mu$SR studies \cite{98_nachumi,17_guguchia}.

The incommensurate peaks of the parallel spin stripe order are clearly observed above $T_\text{2}$ for both Zn-0\% and 2\% samples as seen in Fig. \ref{Fig3}(b) and (c), and order parameter intensity is observed up until $T_\text{1} \sim$ 68\,K for both samples, before its slope goes to zero. The corresponding charge stripe order in Nd-LSCO ($x$ = 0.125) is known to onset below $\sim$ 68\,K as well \cite{95_tranquada,96_tranquada,97_tranquada,01_takeda,21_gupta}.  Our present results clearly show that both spin and charge stripes onset at {\it the same} temperature.

The elastic line scans in Fig. \ref{Fig3}(b) and (c) have been fit to Lorentzian line-shapes to extract widths related to correlation lengths in the $ab$ plane as a function of temperature.  The full widths at half maxima (FWHM) so-extracted are plotted in Fig. \ref{Fig3}(d), and both show a linearly decreasing correlation length (increasing FWHM) at temperatures below $T_\text{2} \sim$ 33\,K and 24\,K for the Zn-0\% and 2\% crystals, respectively. Above $T_\text{2}$, the slope of the decreasing correlation lengths with increasing temperature increases.  This indicates two things: the parallel spin stripe phases in Zn-0\% and 2\% doped Nd-LSCO ($x$ = 0.125) samples do not display true long range order down to 1.4\,K, consistent with earlier results \cite{99_tranquada}; and a transition of sorts occurs at $T_\text{2} \approx$ 33\,K (24\,K) for the Zn-0\% (Zn-2\%) samples.
The transition at $T_\text{2}$ may simply be where the ordered moment participating in the parallel spin stripe structure begins to saturate.  As mentioned above, $\mu$SR, a local probe, detects moments which are static on the muon timescale at $T_\text{2}$ in pure Nd-LSCO ($x$ = 0.125) \cite{98_nachumi,17_guguchia}. But finite correlation lengths within the basal $ab$ plane exist much below $T_\text{2}$, while the correlation length along $c$ is short for all temperatures above 5\,K \cite{96_tranquada}. Hence $T_\text{2}$ may be a crossover temperature scale on which the low energy spin dynamics rapidly evolve and pass through the $\mu$SR time window.

\begin{figure}[!tbp]
\includegraphics[width=6cm]{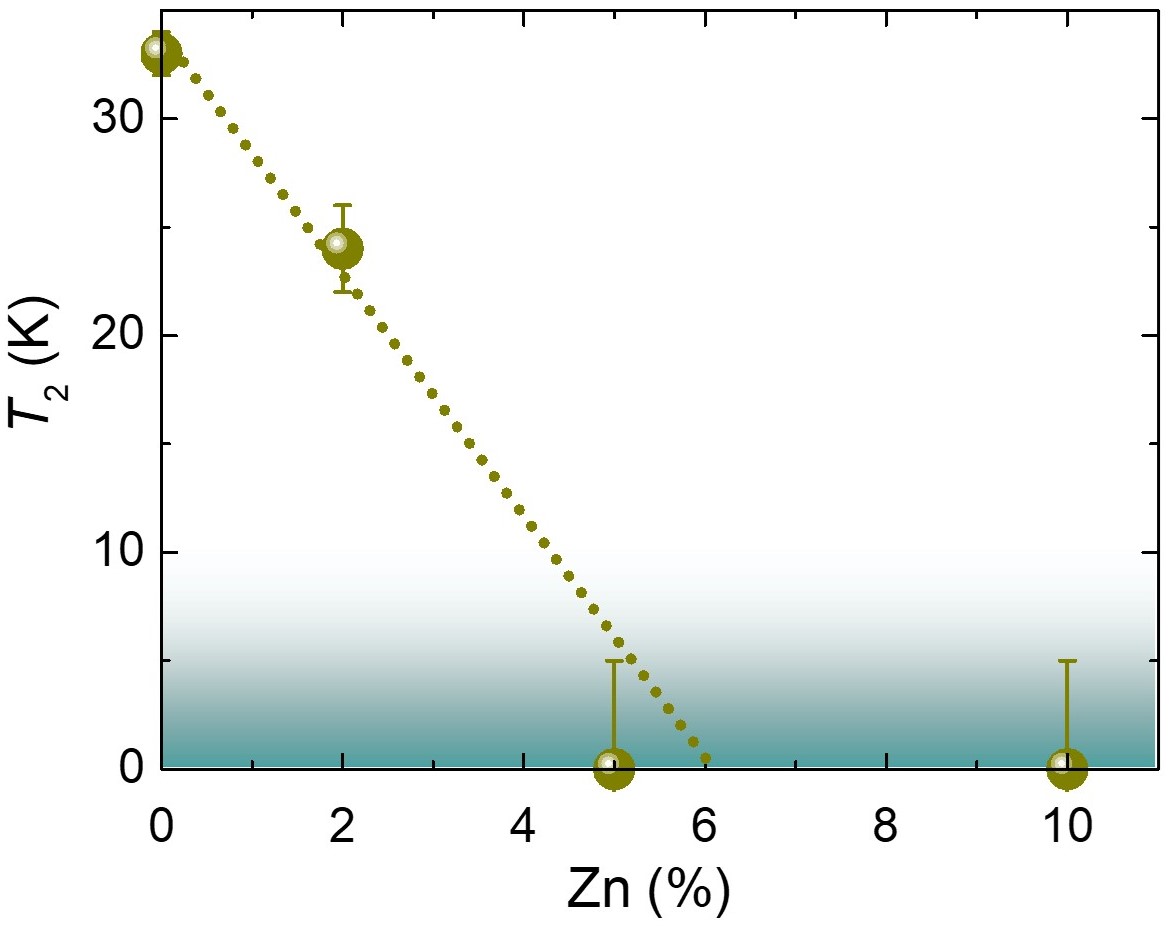}
\caption{The spin stripe transition temperature $T_\text{2}$ for Nd-LSCO ($x$ = 0.125) as a function of Zn concentration, as identified by elastic neutron scattering order parameter measurements.  The shaded temperature region below $\sim$ 7\,K indicates temperatures below which coupling between Nd$^{3+}$ and Cu$^{2+}$ moments strongly influences the stripe structure.
}
\label{Fig4}
\end{figure}

For Zn-5\% and 10\% single crystals, as will be discussed, the much-weaker order parameters at high Zn-doping show upwards curvature at all temperatures above 1.4\,K, hence consistent with $T_\text{2}\sim$ 0.
The $T_\text{2}$'s extracted from this analysis on all four Nd-LSCO single crystal samples are shown in Fig. \ref{Fig4}, and the extreme sensitivity of the parallel spin stripe phase to quenched magnetic vacancies is evident as $T_\text{2}$ appears to be suppressed to zero by Zn-6\%, almost a factor of 7 below the 2D percolation threshold of $\sim$ 41\%.

\begin{figure}[!tbp]
\includegraphics[width=8cm]{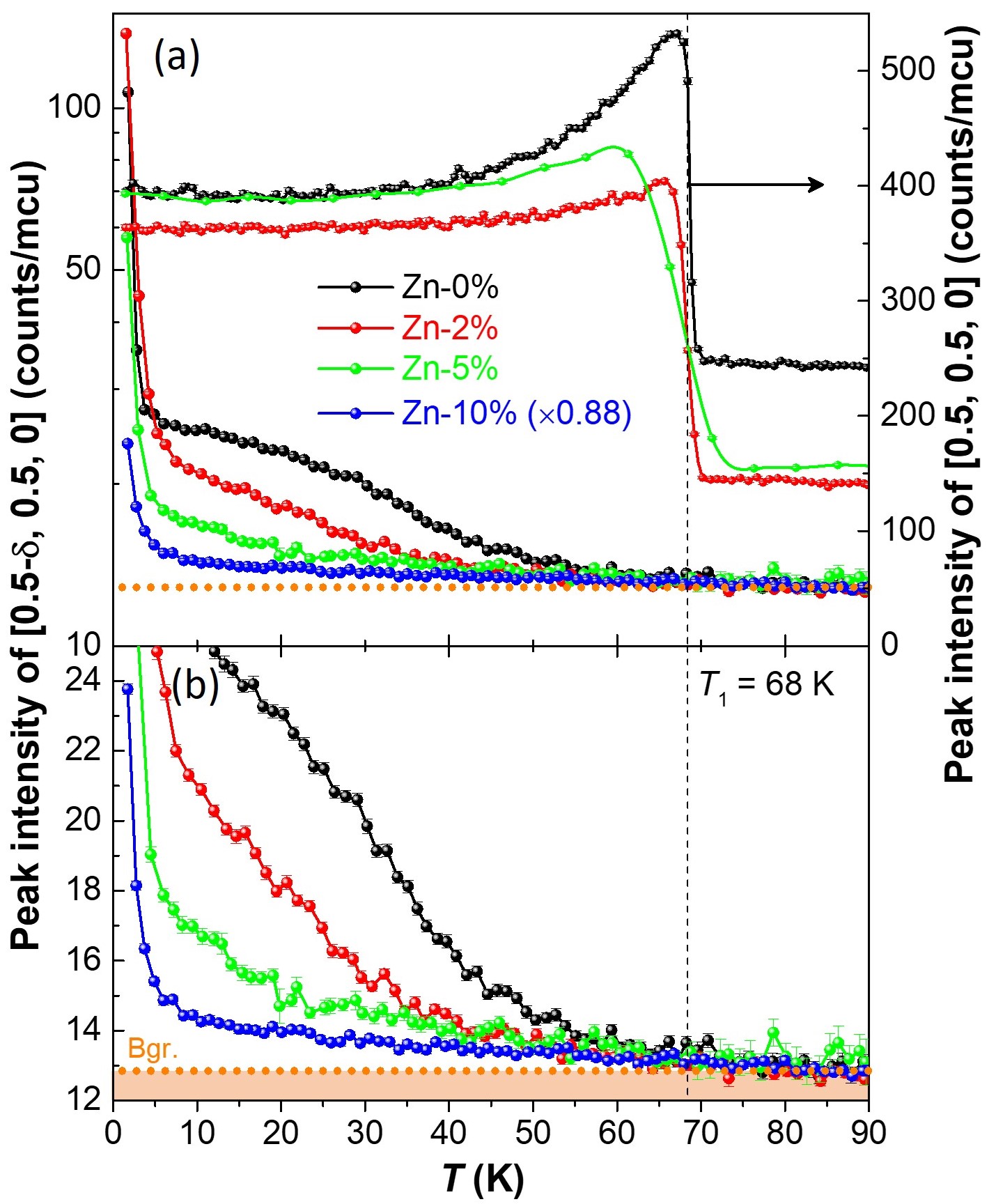}
\caption{The order parameters of Nd-LSCO ($x$ = 0.125) single crystals with four different Zn-doings on both log and linear intensity scales.  The intensity of the Zn-10\% sample has been multiplied by 0.88 in order to have the intensity for all four samples agree for $T > T_\text{1}$, but otherwise the intensities are normalized only to the beam monitor count unit (mcu).  Panel (a) also shows the temperature dependence of the commensurate nuclear Bragg peak at [0.5, 0.5, 0].
}
\label{Fig5}
\end{figure}

A previous $\mu$SR and neutron study by Guguchia {\it et al} \cite{17_guguchia} on La214 cuprate systems showed that the spin stripe order temperature, $T_\text{so}$, responded similarly to Zn-doping as that of $T_\text{2}$ which we obtain by neutron scattering, shown in Fig. \ref{Fig4}. However, the only neutron diffraction data reported in that earlier study, the order parameter of Nd-LSCO ($x$ = 0.125) with Zn-1.6\% doping, shows this order parameter goes to zero at $T_\text{so} \approx$ 10\,K, a result which is inconsistent with our order parameter for Nd-LSCO ($x$ = 0.125) with Zn-2\% in Figs. \ref{Fig3}(f), \ref{Fig4}, and \ref{Fig5}, and with our [$0.5- \delta$, 0.5, 0] line scans above 10\,K for the same Zn-2\% sample as shown in Fig. \ref{Fig3}(c). Nonetheless, our present elastic neutron scattering results and the Guguchia {\it et al} $\mu$SR results lead to very similar and striking sensitivity of parallel spin stripe phase to quenched nonmagnetic disorder.  The Guguchia {\it et al} work also showed that superconducting $T_\text{c}$ possesses a similar Zn-doping sensitivity for LSCO and LBCO near optimal doping for superconductivity.

We then summarize the full order parameters of all four Zn-doped Nd-LSCO ($x$ = 0.125) single crystals in Fig. \ref{Fig5} on both log and linear intensity scales, respectively. Panel (a) also shows the temperature dependence of the commensurate nuclear Bragg peak at [0.5, 0.5, 0]. Its abrupt drop in intensity at $T_\text{1}$ signals the LTT to LTO structural phase transition, which broadens slightly with increasing Zn-doping, but does not significantly move in temperature. The parallel spin stripe order parameters are severely and systematically affected by the Zn-doping from Zn-2\% to 10\%, despite the fact that they are all well below the 2D percolation threshold of $\sim 41\%$.  At 5\% and 10\% Zn-doping, the peak intensity of the parallel spin stripe quasi-Bragg peak is sufficiently weak that a well defined peak, similar to what is shown in Fig. \ref{Fig3}(b-c), is not easily identifiable above 10\,K. As discussed, the Zn-5\% and Zn-10\% order parameters show upwards curvatures at all temperatures.

This is very reminiscent of the order parameter for CsCo$_{0.83}$Mg$_{0.17}$Br$_3$, in the temperature regime for the partially-paramagnetic N\'eel state displayed by pure CsCoBr$_3$. It was attributed to the quenched impurities coupling to the inhomogeneous N\'eel state as a random field \cite{04_duijn}. A similar interpretation is relevant here, again due to an inhomogeneous ordered state - in this case the parallel stripe phase.

This is an interesting conclusion for at least three reasons. First, it presents a rare-example of a systematic study of random field effects from quenched disorder in an inhomogeneous ordered magnetic state. Second, it provides strong evidence that the incommensurate, ordered structure below $\sim$ 50\,K in Nd-LSCO ($x$ = 0.125) and related La214 cuprates is an {\it inhomogeneous} parallel spin stripe phase and not a {\it homogeneous} spiral SDW. Third, and perhaps most  importantly, it implies that quenched non-magnetic impurities are preferentially located coincident with the parallel charge stripe component of the parallel stripe structure, as the parallel charge stripes will seek them out to lower the magnetic energy of the inhomogeneous parallel stripe structure. Quenched magnetic vacancies can therefore be very effective at breaking up Cooper pairs propagating along the charge stripes, and can thereby account for the extreme sensitivity of the superconducting $T_\text{c}$ to non-magnetic disorder in the CuO$_2$ plane in Nd-LSCO ($x$ = 0.125), and by extension in other cuprate superconductors for which an inhomogeneous stripe phase is relevant.

This work was supported by the Natural Sciences and Engineering Research Council of Canada. A portion of this research used resources at the High Flux Isotope Reactor, a DOE Office of Science User Facility operated by the Oak Ridge National Laboratory.  We acknowledge stimulating discussions with A.-M. S. Tremblay, A. Sacuto, E. S. S{\o}rensen, and A. D. S. Richards.

\bibliography{NdLSCO_Zn}

\end{document}